\documentclass[pra,twocolumn,aps,showpacs]{revtex4-1}
\usepackage{graphicx}
\usepackage{physics}
\newcommand{\etal}{\textit{et al}.~}
\newcommand{\ie}{\textit{i}.\textit{e}.}

\usepackage{xcolor}
\definecolor{maroon}{RGB}{100,20,20}
\definecolor{dblue}{RGB}{20,20,100}
\usepackage[colorlinks=true,linkcolor=dblue,citecolor=maroon,urlcolor=maroon]{hyperref}
\begin{document}
\title{Do weak  values capture the complete truth about the
past of a quantum particle?}
\author{Rajendra Singh Bhati}
\email{ph16076@iisermohali.ac.in}
\affiliation{Department of Physical Sciences, Indian
Institute of Science Education and
Research (IISER) Mohali, Sector 81 SAS Nagar, Manauli PO
140306 Punjab India}
\author{Arvind}
\email{arvind@iisermohali.ac.in}
\affiliation{Department of Physical Sciences, Indian
Institute of Science Education and
Research (IISER) Mohali, Sector 81 SAS Nagar, Manauli PO
140306 Punjab India}
\begin{abstract}
Weak values inferred from weak measurements have been
proposed as a tool to investigate trajectories of pre- and
post-selected quantum systems.  Are the inferences drawn
from the weak values about the past of a quantum particle
fully true? Can the two-state vector formalism predict
everything that the standard formalism of quantum mechanics
can?  To investigate these questions we present a
``which-path'' gedanken experiment in which the information
revealed by a pre- and post-selected quantum system is
surprisingly different from what one would expect from the
weak values computed using the two-state vector formalism.
In our gedanken experiment, a particle reveals its presence
in locations where the weak value  of the projection
operator onto those locations was vanishingly small.
Therefore our predictions turn out to be in
contradistinction to those made  based on  the non-vanishing
weak values as the presence indicators of the quantum
particle. We propose a six port photon-based interferometer
setup as a possible physical realization of our gedanken
experiment.  
\end{abstract} 

\maketitle
\section{Introduction}
The value of an observable does not hold a meaning prior to
the measurement in the standard formalism of quantum
mechanics~\cite{BOHR1928,Wheeler19789,
Heisenberg1958-HEIPAP,Scully1991}; however, in the
time-symmetric two-state vector formalism (TSVF) of quantum
mechanics such a meaning is alluded to via ``weak
values''~\cite{PhysRevLett.60.1351,
PhysRevA.41.11,0305-4470-24-10-018}.
TSVF and the concept of weak values were introduced to
validate the retrodiction formula introduced by Aharonov,
Bergmann and Lebowitz (ABL rule) to calculate probabilities
of counterfactual-measurement outcomes of an observable for
a pre- and post selected ensemble~\cite{PhysRev.134.B1410}.
In TSVF, the weak values fully determine the
properties of pre- and post-selected quantum system at all
intermediate times.  The weak values can lie outside the
range of the values of an observable allowed by standard
quantum mechanics and they need not even be real.

Experimental observations of weak values have been
carried out
~\cite{PhysRevLett.66.1107,PhysRevLett.94.220405,
PhysRevLett.100.056801}, even in the absence of a consensus
on their interpretation~\cite{PhysRevLett.62.2325,
PhysRevLett.113.120404,SOKOLOVSKI2018382,SOKOLOVSKI20161593}.
The concept of weak values has been used in understanding
optical telecom networks ~\cite{PhysRevLett.91.180402},
superluminal and slow light phenomena in birefringent
photonic crystal
~\cite{PhysRevLett.92.043601,PhysRevLett.93.203902},
studying optical cross-phase modulation
jump~\cite{PhysRevLett.102.013902}, quantum process
tomography~\cite{Kim2018}, ultrasensitive quantum
measurements using weak value
amplification~\cite{Hallaji2017,PhysRevLett.112.200401,
PhysRevLett.107.133603}, and a review is available in
reference~\cite{RevModPhys.86.307}.  Weak values have also
been used in direct measurements of wavefunctions and in
providing an operational definition to the
wavefunction~\cite{Lundeen2011,PhysRevLett.108.070402}.

Apart from their applications, weak values have been thought
to provide insights into a number of fundamental issues in
quantum mechanics, which include Hardy's
paradox~\cite{PhysRevLett.68.2981,AHARONOV2002130,PhysRevLett.95.030401,
PhysRevLett.102.020404}, quantum tunneling
time~\cite{PhysRevLett.74.2405}, the Legget-Garg
inequality~\cite{PhysRevLett.100.026804}, Bohmian
trajectories~\cite{Kocsis1170,Mahlere1501466} and quantum
contextuality~\cite{PhysRevLett.113.200401}.  These studies
are firmly based on the straightforward interpretation that
{\it the weak value is the value of an observable
between two successive measurements} of a quantum system.
The weak values have also  been called the weak-measurement
elements of reality (WMER)~\cite{Vaidman1996} and  it has
been proposed that the trace a particle leaves at a location
is proportional to the weak value of the projection operator
onto that particular location~\cite{PhysRevA.87.052104}.

The use of straightforward interpretation of weak value in
few experimental schemes has resulted in inception of new
quantum paradoxes: the paradox of negative number of
particles and negative
pressure~\cite{0305-4470-24-10-018,RESCH2004125}, the
paradox of discontinuous trajectories of a
photon~\cite{PhysRevLett.98.160403,PhysRevA.87.052104,
PhysRevLett.111.240402,PhysRevA.95.042121,Aharonov2017,
deLimaBernardo2017}, and the paradox of quantum Cheshire
cat~\cite{1367-2630-15-11-113015,Denkmayr2014,Das_2020,Liu2020}.
The last two have been at the center stage of the discussion
for researchers working on quantum foundations. The most
surprising and `common sense' defying claim made by Danan
\etal~\cite{PhysRevLett.111.240402} is that {\it a pre- and
post-selected photon in a nested Mach-Zehnder
interferometer} (NMZI) {\it takes discontinuous trajectories
to reach the detector}. The photon visits a region in the
NMZI without entering and exiting it. Another `common sense'
defying claim is made by Aharonov
\etal~\cite{1367-2630-15-11-113015} and Denkmayr
\etal~\cite{Denkmayr2014} that {\it the internal degree of
freedom of a quantum system can be separated from its
wavefunction.} Many comments and papers have been published
in criticism as well as defense of these
claims~\cite{PhysRevA.96.022126,1751-8121-49-34-345302,
Alonso2015,PhysRevA.94.032115,PhysRevA.97.052111,WIESNIAK20182565,
Atherton:15,1367-2630-17-5-053042,DUPREY20181}.

In this paper, using a gedanken experiment
involving time varying Hamiltonians, we show that weak values do
not capture the complete truth about the past of a quantum
particle. This is in contradistinction to the claim that
TSVF is a complete description of a pre- and post-selected
quantum system. The signatures of the Hamiltonian evolution
and the time dependence of the Hamiltonian which are present
in the quantum state just before measurement, are typically
lost in the postselection process for the individual
particle. However, these signatures can be present in the
probability distribution measured over time. In our 
gedanken
experiment, we use this technique to investigate the past of
a post-selected quantum system. Certain time dependent
elements in the Hamiltonian at certain locations which
oscillate at fixed frequencies are carefully inserted. We
then use the presence of these frequencies in the measured
probabilities as indicators for the passage of the particle
through locations where such time dependent elements were
installed. In our analysis we arrive at a conclusion that
the weak values of the TSVF formalism at times fail to
capture the presence of the quantum particles.

The material in this paper is arranged as
follows:~Section~\ref{ABL_background} gives a brief review
of Aharonov-Bergmann-Lebowitz (ABL) rule, TSVF, Weak Values
and their interconnections. Section~\ref{paradoxes} gives
background of leading weak value paradoxes.
Section~\ref{gedanken} describes our gedanken experiment
aimed at providing a counterexample to the weak value based
interpretation of the past of a quantum particle.  In
Section~\ref{TSVF} we compare the predictions of our
nalysis with those of TSVF to demonstrate the  mismatch. In
section~\ref{CH_analysis} we provide a consistent histories (CH)
analysis of our six-port interferometer.  Section~\ref{conc}
provides conclusions and discussion.
\section{ABL Rule, TSVF, Weak Values and Weak Value Paradoxes}
\label{ABL_background}
According to the
ABL rule, the measurement of an observable $A$ of a quantum system
at time $t$ which is pre-selected in state $\ket{\psi_{1}}$
at time $t_1<t$ and post-selected in state $\bra{\psi_{2}}$
at time $t_2>t$ would yield eigenvalue $a_n$ with
probability~\cite{PhysRev.134.B1410,Aharonov_1991}:
\begin{equation}
\label{ABL_rule_TSVF_degenerate}
P_{t}(a_n|\psi_1,\psi_2)=\frac{|\bra{\psi_2(t)}\Pi_{a_n}\ket{\psi_1(t)}|^{2}}
{\sum_{i}{|\bra{\psi_2(t)}\Pi_{a_i}\ket{\psi_1(t)}|^{2}}}
\end{equation}
where
$\Pi_{a_i}=\sum_{\alpha}\ket{a_{i,\alpha}}\bra{a_{i,\alpha}}$
with $\{\ket{a_{i,\alpha}}\}$ being a complete set of
eigenstates of $A$ labeled by eigenvalues $a_i$, and
$\ket{\psi_j(t)}=\exp[-\frac{i}{\hbar}
\int_{t_j}^{t}Hdt]\ket{\psi_j}$ with $j=1,2$.
It can be seen that the state $\ket{\psi_1(t)}$ evolves forward while
the state $\ket{\psi_2(t)}$ evolves backward in time both being on equal
footing in the TSVF formalism.
Application of ABL rule in counterfactual reasoning results
in surprising and paradoxical situations. For example
consider the three box problem~\cite{Aharonov_1991}: given that a
particle is prepared in superposition of being in three
non-overlapping boxes $A, B$ and $C$ with state
$\ket{\psi_1}=\frac{1}{\sqrt{3}}(\ket{A}+\ket{B}+\ket{C})$
and post-selected in state
$\ket{\psi_2}=\frac{1}{\sqrt{3}}(\ket{A}+ \ket{B}-\ket{C})$,
the probability of finding the particle in box $A$ or $B$ upon
opening the respective box at any intermediate time is one
according to the ABL rule. In other words if either of the boxes
$A$ and $B$ had been opened at an intermediate time, one
would always find the particle there.

Application of the ABL rule in time symmetric counterfactual
reasoning faced a serious refutation from
Kastner~\cite{Kastner1999}, Miller~\cite{MILLER199631},
Cohen~\cite{PhysRevA.51.4373} and others on
philosophical grounds. Giving an alternative interpretation
of the ABL rule as being the probability of the outcome of an
actual measurement of the observable between pre-selection
and post-selection measurements, these authors pinpointed that
the paradoxes arise only when one uses the ABL rule for
calculating probabilities of possible outcomes of
observables which actually have not been measured at
intermediate times. While the
debate about the interpretation of
the ABL rule was thought to be settled with experimental
realizations of counterfactual paradoxes using weak values
introduced as a witness for the ABL rule, as we shall
see the predictions of the ABL rule remains questionable.
According to
the measurement postulate of quantum theory, performing an
actual measurement on a quantum system at the intermediate
time would destroy the prepared state making counterfactual
interpretation no longer valid. Then the question is how to
experimentally witness the counterfactualness of the ABL rule.
The answer is that the ABL probabilities given by
Equation~(\ref{ABL_rule_TSVF_degenerate}) can be
inferred using weak values 
$\Pi_{a_i}^{w}(t)$
of the projection operators
\{$\Pi_{a_i}$\} at the intermediate time $t$, given as:
\begin{equation} \label{WV}
\Pi_{a_i}^{w}(t)
=\frac{\bra{\psi_2(t)}\Pi_{a_i}
\ket{\psi_1(t)}}{\bra{\psi_2(t)}\ket{\psi_1(t)}}
\end{equation}
which can be experimentally determined without collapsing
the wave function.  The concept of weak values, has thus
been claimed to have the potential to provide a ground for an
experimental realization of the ABL rule without performing
a projective  measurement at intermediate times and thereby
giving the counterfactual interpretation an operational
meaning.

Let us for a moment revisit, the three box problem. If any
of the boxes $A$ and $B$ had been opened, according to
counterfactual ABL rule, the particle would have been found
with certainty (probability one) in the respective box.
This raises a serious and natural question: how can a single
particle be present in more than one boxes with certainty?
The concept of weak values resolve this problem. It has been
hypothesized that, \textit{the
weak values are values of corresponding observables and
fulfill the conditions of being elements of reality of weak
measurements (WMER)}~\cite{Vaidman1996, Aharonov_1991}.  Let
us call it the \textit{weak value hypothesis} (WVH). The
validity of WVH naturally leads one to conclude that the
weak value of a projection operator $\ket{\eta} \bra{\eta}$
is the number of quantum systems present in the state
$\ket{\eta}$.  Therefore, the number of particles present in
boxes $A, B$ and $C$ are $1, 1$ and $-1$ respectively
keeping the total number of particles one at any
intermediate time. As one can see one has to accept the
concept of negative number particles in this explanation!

A natural consequence of WVH is the truthfulness of
the following statement:

S-$A$: \textit{If the weak value of the projection
operator $\Pi_{x}=\ket{x}\bra{x}$ at an intermediate time is
zero, where $\ket{x}$ is a position eigenstate; then the
particle was not present at position $x$ at that time.}

The above statement is just a codification of the counterfactual
statement: that if $P_{t}(a_n|\psi_1,\psi_2)=0$ then the
measurement of observable $A$ on system at the intermediate
state would never yield value $a_n$.  Since, the whole
purpose of bringing the concept of weak measurements was to
provide an operational meaning to counterfactual ABL rule in
terms of weak values, one can write an operational
definition of the past of a quantum particle, as has been
done by Vaidman, using the concept of weak values:

S-$B$: \textit{A quantum particle was
present at a location if and only if it left a weak trace on
a pointer located at that location upon interaction.}

The weak traces can be experimentally measured by a complete
state tomography of the pointer state after the
post-selection of the system state. Since the
post-selection measurement leaves the state of the system
and the pointer separable, a further measurement on
the pointer
state will definitely not affect the past of the system in a
retro-causality manner.

%


\section{Weak value paradoxes}
\label{paradoxes}
We describe here situations  where WVH  leads to paradoxical
situations. Therefore, a closer examination of the WVH
and of experiments which may contradict this hyposthesis is 
important and is the focus of our work.
\subsection{Photon with discontinuous trajectories}
\begin{figure}
\includegraphics[scale=1]{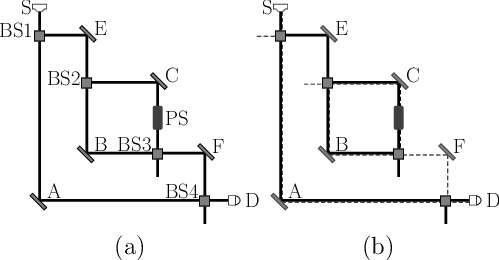}
\caption{
(a) Setup of a nested Mach-Zehnder interferometer. S
is a single
photon source, D is a photon detector, PS is a
phase shifter while BS1
and BS4 are beamsplitters with $2/3$ reflectivity and  BS2 and
BS3 are 50-50 beamsplitters. (b) Solid and dashed lines
represent forward and backward evolving state vectors
respectively.}
\label{NMZI}
\end{figure}

An intriguing example of the weak value paradox is the past of a
photon in a nested Mach-Zehnder interferometer (NMZI), as
investigated by Lev Vaidman~\cite{PhysRevLett.98.160403,
PhysRevA.87.052104}, where photons take discontinuous
trajectories to reach the detector. The experimental setup
is shown in Figure~\ref{NMZI}. A single photon is pre-selected at
source S and post-selected at detector D. If the phase
shifter (PS) is tuned
in such a way that there is a completely destructive
interference near mirror F, then the weak values of
projection operators near mirrors A, B, C, E, and F are
$\Pi^w_A=1,\Pi^w_B=1,\Pi^w_C=-1,\Pi^w_E=0$, and $\Pi^w_F=0$
respectively. Therefore according to S-$A$ the photon
was never present near mirror E and F but it was present
near mirror A, B and C leading us to conclude that the
photon took discontinuous trajectories to reach the detector D.
In the language of counterfactual ABL rule, if one had
placed a position measurement device near mirror E and F to
find whether the photon passed through the regions, one
would have never found a photon.

The experiments of Danan
\etal~\cite{PhysRevLett.111.240402} and Zhou
\etal~\cite{PhysRevA.95.042121} were aimed at experimental
realisation of Vaidman's predictions using weak measurements
in the context of NMZI.
Vaidman's claims and the experimental results of Danan \etal
have faced serious criticism
including from  Englert
\etal~\cite{PhysRevA.96.022126} using `unambiguous
which-path information' (UWI), and R. B.
Griffiths~\cite{PhysRevA.94.032115} using the consistent
histories approach argued that the paradox of
discontinuous trajectories arises from discarding second and
higher order perturbation terms in the interaction strength
$\kappa$. Li \etal \cite{PhysRevA.88.046102} and  D.
Sokolovski~\cite{SOKOLOVSKI2017227} have also been critical
of the TSVF interpretation and they too rely on arguments
based on neglecting the higher order terms. The critique of
the TSVF interpretation of the Danan experiment by 
these authors is based primarily on discarding certain higher order terms.
This we think is insufficient to reject
the main claims because an experimental realization of
the counterfactual ABL rule requires negligible disturbance to the system, a limit in
which the contribution of the higher order terms may be
ignorable. According to counterfactual ABL rule, in
the absence of a weak measurement
device, 
(coupling constant $\kappa= 0$) 
no photon would have been detected if a
photon detector were placed near E or F.
However, in the presence of a weak measuring device
($\kappa\neq 0$) a fraction proportional to
$\kappa^2$ of the pre- and
post-selected ensemble would have been detected near E and F. 
Griffiths has also agreed to this point
in his consistent histories analysis of Danan
experiment~\cite{PhysRevA.94.032115}. As we will see, while
not disagreeing with these observations, we intend to
analyse the situation from a different point of view and
show that the conclusions based on the ABL rule can be in contradiction
to those based on standard quantum mechanics in a concrete 
experimental NMZI.

\subsection{Quantum Cheshire Cat}
\begin{figure}
\includegraphics[scale=1]{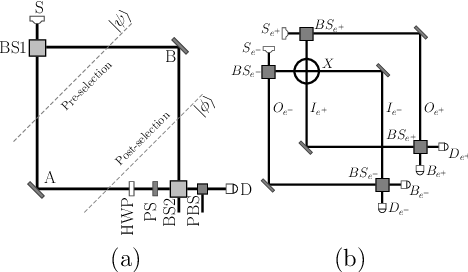}
\caption{
(a) Experimental setup for photonic quantum Cheshire cat.
(b) Hardy's setup: $e^{\pm}$ are created in sources
$S_e^{\pm}$, detected in
detectors $D_e^{\pm}$ and $B_e^{\pm}$. $BS_e^{\pm}$ are
50-50 beam-splitters of respective particles.}
\label{C_H}
\end{figure}

Aharonov \etal ~\cite{1367-2630-15-11-113015} predicted that
the internal degree of freedom of a photon (grin) can be
separated from its wavefunction (cat) which has been
experimentally supported by Denkmayr
\etal~\cite{Denkmayr2014} and Ashby \etal
~\cite{PhysRevA.94.012102}.  The claims are firmly based on
truthfulness of S-$A$. The proposed experimental
setup, shown in Figure~\ref{C_H}(a), is a modified
Mach-Zehnder interferometer in which a photon source S and
a beam-splitter BS1 are used to preselect a single photon in
the state $\ket{\psi}= (\ket{A}+i\ket{B})\ket{H}/\sqrt{2}$ and
a half wave plate (HWP), a phase shifter (PS), a beam splitter
(BS2), a polarizing beam splitter (PBS) and a single photon
detector (D) are used to post-select the photon in the state
$\ket{\phi}=(\ket{A}\ket{V}+ \ket{B}\ket{H})/\sqrt{2}$. Here
states $\ket{A}$ and $\ket{B}$ are spatial state vectors of
photon being in arm A and B respectively and, $\ket{H}$ and
$\ket{V}$ are horizontal and vertical polarization states
respectively. States with circular polarization
are $\ket{\pm}= (\ket{H}\pm i\ket{V})/\sqrt{2}$. Let us
now ask a question: which arm did photon pass through to
reach detector D and which of the two ciruclar polarisation
(determined by measureing $\sigma_z=\ket{+}\bra{+}
-\ket{-}\bra{-}$) it was possessing in each arm?  
To answer the question we calculate weak values of
operators $\Pi_A=\ket{A}\bra{A}, \Pi_B=\ket{B}\bra{B},
\sigma_z^{A}=\Pi_A\otimes\sigma_z$ and $\sigma_z^{B}
=\Pi_B\otimes\sigma_z$ as:
\begin{equation}
\label{cheshire_VW}
\begin{aligned}
(\Pi_A)^w &=0; & (\Pi_B)^w =1\\
(\sigma_z^{A})^w & =1; & (\sigma_z^{B})^w =0 \\
\end{aligned}
\end{equation}
VWH implies that the circular polarization of the photon in arm A
was non-zero but it did not pass through arm A while photon
passed through arm B but the circular polarization in that
arm was zero. This is how Aharonov \etal 
concluded that
photon wavefunction (cat) was separated from polarization
(grin).
\subsection{Weak value version of Hardy's Paradox}
Hardy~\cite{PhysRevLett.68.2981} used a gedanken
experiment involving a bipartite system to
give a logical proof against local realism 
and to show a contradiction between quantum mechanics and any
realistic theory which has Lorentz invariant elements-of-reality.
The paradox arises when one uses counterfactual reasoning
about the past of a system in Hardy's
setup~\cite{AHARONOV2002130}.
The setup consists of two Mach-Zehnder
interferometers as shown in Figure~\ref{C_H}(b), one
($I_{e^-}$) for an
electron and the other ($I_{e^+}$) for a simultaneously
produced positron.  X is an overlapping region of inner arms
of both the interferometers ($I_{e^{\pm}}$) such that if
positron and electron encounter each other
the annihilation occurs
with probability one.  Arms of the two interferometers are adjusted
in such a way that there is no detection in $D_{e^{\pm}}$
when the two interferometers are separated 
so that there
is no overlapping region.  When there is an overlap of
$I_{e^{\pm}}$, $e^{-}-e^{+}$ annihilation in region X act as
a Elutzer-Vaidman bomb~\cite{Elitzur1993} and disturbs the
interference causing coincident detection in $D_{e^{\pm}}$
with probability $1/16$. Let us now ask a question: which
arms did $e^+$ and $e^-$
travel through when
there was a coincidence count \ie both  $D_{e^{\pm}}$ detected a particle simultaneously? A counterfactual
reasoning leads to the following paradox: if there is a
coincident detection in $D_{e^{\pm}}$ then $e^{\pm}$ must
have travelled through region X in order to disturb the
interference, but there is no annihilation! 
Aharonov \etal ~\cite{AHARONOV2002130} made
the paradox even weirder when they use the ABL rule and
VWH.
They conclude:
{\it 
(a): $e^+$ always passed through
region X. (b): 
$e^-$ always passed through region X.
(c):
$e^-$ and $e^+$ 
together never passed through the region X.
 } 
While there have been experiments on the weak value
version  
of Hardy's paradox
based on the measurement of weak
values~\cite{PhysRevLett.102.020404,
1367-2630-11-3-033011}, the interpretation requires one to 
accept VWH. Therefore, it is important to scrutinise  
VWH further.
\section{The gedanken experiment}
\label{gedanken}
We now describe our main results where we will consider
a situation where the predictions of weak values can be
seen to come in contradition with descriptions based on standard
quantum mechanics. Consiser 
A quantum system with a six-dimensional Hilbert space
$\mathcal H$.  For the purpose of the gedanken
experiment, we can think of a quantum particle being in six
non-overlapping boxes.  If the particle is found  in the
$i^{th}$ box with certainty, the state vector of the
particle is written as $\ket{i}$. In the absence of
interactions these states are orthogonal to each other.  The
boxes are designed in such a way that the interactions can
be switched on so that the particle can tunnel between any
pair of boxes in a controlled manner.  The boxes $i$ and $j$
can be made to interact instantaneously at time $t^{\prime}$
via the interaction Hamiltonian
$H^\prime=g\delta(t-t^\prime)\sigma_{y}^{(ij)}$.  Here
$\sigma_{y}^{(ij)}=i(-|{i}\rangle\langle{j}|+|{j}\rangle\langle{i}|)$
is $\sigma_{y}$ Pauli matrix and $\delta(t-t^\prime)$ is a
Dirac delta function of time $t$. The tunable parameter $g$
represents the tunneling strength and we call the process as
a {\it leakage} process when $g$ is sufficiently small.
Further,  
the \textit{operational condition} $g^2N\approx{1}$ has to
be satisfied where
$N$ a large number representing ensemble size being  considered
by the experimenter. Therefore,  we need to retain only
the terms linear in $g$ unless it is multiplied by $N$. In
the rest of the paper, whenever we neglect the contribution
of  higher powers of some quantity, it is understood that we
are assuming that the \textit{operational condition} is
satisfied.

Consider a quantum state $\ket{\psi(t_0)}$ of the
particle at $t_0$ which
undergoes  time evolution according to the Hamiltonian:
\begin{eqnarray}
\nonumber 
H&=&\sum^{9}_{i=1}H_{i}, \quad\quad {\rm With}\\
\nonumber 
H_{1} & = & -\hbar\{\sin^{-1}({\sqrt{{2}/{3}}})
\delta(t-t_{1})(\sigma_{y}^{(13)}+\sigma_{y}^{(24)})\} \\
\nonumber 
H_{2} & = &
-\epsilon\hbar\cos{(\omega_{1}t)}\delta(t-t_{2})\sigma_{y}^{(34)} \\
\nonumber 
H_{3} & = & -\frac{\pi}{4}\hbar\delta(t-t_{3})(\sigma_{y}^{(35)}+
\sigma_{y}^{(46)}) \\
\nonumber 
H_{4} & = & -\epsilon\hbar\delta(t-t_{4})
\{\cos{(\omega_{2}t)}\sigma_{y}^{(12)}+\cos{(\omega_{3}t)}\sigma_{y}^{(34)} \\ 
\nonumber 
& & +\cos{(\omega_{4}t)}\sigma_{y}^{(56)}\} \\
\nonumber 
H_{5} & = &
-\frac{\pi}{2}\hbar\delta(t-t_{5})(I^{(56)}-\sigma_{z}^{(56)}) \\
\nonumber  
H_{6} & = & -\frac{\pi}{4}\hbar\delta(t-t_{6})
(\sigma_{y}^{(35)}+\sigma_{y}^{(46)}) \\
\nonumber 
H_{7} & = & -\epsilon\hbar\cos{(\omega_{5}t)}
\delta(t-t_{7})\sigma_{y}^{(34)} \\
\nonumber 
H_{8} & = & -\hbar\{\sin^{-1}{(\sqrt{{2}/{3}})}
\delta(t-t_{8})(\sigma_{y}^{(13)}+\sigma_{y}^{(24)})\} \\
H_{9} & = & -\frac{\pi}{4}\hbar\delta(t-t_{9})\sigma_{y}^{(12)}
\label{Hamiltonian}
\end{eqnarray}
where $I^{(ij)}=|{i}\rangle\langle{i}|+|{j}\rangle\langle{j}|$ and
$\sigma_{z}^{(ij)}=|{i}\rangle\langle{i}|-|{j}\rangle\langle{j}|$.
The impulsive interaction occurs at 
moments of time
$t_0<t_1<t_3<{\cdots}<t_8<t_{9}$. The
parameter 
$\epsilon\ll{1}$ is such that the contributions of higher
powers of $\epsilon$ in the experimental observations are
negligible. Therefore, $H_2,H_4,$ and $H_7$ generate
\textit{leakage} processes between certain boxes. 
The time intervals between
$t_i$'s  are kept fixed for repeated runs of the
experiment. Since all the transformations generated by
$H_i$ are momentary and
well separated in time, the state of the particle at time
$t>t_{9}$ is given for infinitesimally small $\Delta$ as
\begin{equation}
\label{Hamiltonian_effective}
\begin{split}
\ket{\psi(t)}=\exp[-\frac{i}{\hbar}\int^{t_{9}+\Delta}_{t_{9}-\Delta}H_{9}dt]
\exp[-\frac{i}{\hbar}\int^{t_{8}+\Delta}_{t_{8}-\Delta}H_{8}dt]{\cdots}
\\ 
{\cdots}\exp[-\frac{i}{\hbar}\int^{t_{2}+\Delta}_{t_{2}-\Delta}H_{2}dt]
\exp[-\frac{i}{\hbar}\int^{t_{1}+\Delta}_{t_{1}-\Delta}H_{1}dt]\ket{\psi(t_{0})}
\end{split}
\end{equation}

The sequence of momentary interactions presented in Equation~(\ref{Hamiltonian})
and time evolution of the system shown in Equation~(\ref{Hamiltonian_effective})
can be understood as a sequence of unitary operations $U_1,U_2,\cdots,U_9$ acting
on the system at times $t_1,t_2,\cdots,t_9$ respectively, where
\begin{equation}
U_{j}=\exp[-\frac{i}{\hbar}\int^{t_{j}+\Delta}_{t_{j}-\Delta}H_{j}dt]
\end{equation}

Unitary operations $\{U_i\}$ are $6\times{6}$ matrices:

\[
U_{1}=U_{8}=\left[
\begin{array}{ccc}
\frac{1}{\sqrt{3}}\mathbf{I} & \sqrt{\frac{2}{3}}\mathbf{I} & 
\mathbf{0} \\
-\sqrt{\frac{2}{3}}\mathbf{I} & \frac{1}{\sqrt{3}}\mathbf{I} & \mathbf{0}  \\ 
\mathbf{0} & \mathbf{0} & \mathbf{I}  
\end{array}
\right];
U_{2}=\left[
\begin{array}{ccc}
\mathbf{I} & \mathbf{0} & \mathbf{0} \\
\mathbf{0} & \mathbf{L_{1}} & \mathbf{0} \\
\mathbf{0} & \mathbf{0} & \mathbf{I}
\end{array}
\right]
\]

\[
U_{3}=U_{6}=\left[
\begin{array}{ccc}
\mathbf{I} & \mathbf{0} & \mathbf{0} \\
\mathbf{0} & \frac{1}{\sqrt{2}}\mathbf{I} & \frac{1}{\sqrt{2}}\mathbf{I}  \\
\mathbf{0} & -\frac{1}{\sqrt{2}}\mathbf{I} & \frac{1}{\sqrt{2}}\mathbf{I}
\end{array}
\right];
U_{4}=\left[
\begin{array}{ccc}
\mathbf{L{2}} & \mathbf{0} & \mathbf{0} \\
\mathbf{0} & \mathbf{L_{3}} & \mathbf{0} \\
\mathbf{0} & \mathbf{0} & \mathbf{L_{4}}
\end{array}
\right]
\]

\[
U_{5}=\left[
\begin{array}{ccc}
\mathbf{I} & \mathbf{0} & \mathbf{0} \\
\mathbf{0} & \mathbf{I} & \mathbf{0} \\
\mathbf{0} & \mathbf{0} & \sigma_{z}
\end{array}
\right];
U_{7}=\left[
\begin{array}{ccc}
\mathbf{I} & \mathbf{0} & \mathbf{0} \\
\mathbf{0} & \mathbf{L_{5}} & \mathbf{0} \\
\mathbf{0} & \mathbf{0} & \mathbf{I}
\end{array}
\right];
U_{9}=\left[
\begin{array}{ccc}
\mathbf{R} & \mathbf{0} & \mathbf{0} \\
\mathbf{0} & \mathbf{I} & \mathbf{0} \\
\mathbf{0} & \mathbf{0} & \mathbf{I}
\end{array}
\right]
\]

where
\[
\mathbf{I}=\left[
\begin{array}{cc}
1 & 0 \\
0 & 1
\end{array}
\right] ;
\mathbf{0}=\left[
\begin{array}{cc}
0 & 0 \\
0 & 0
\end{array}
\right] ;
\mathbf{R}=\frac{1}{\sqrt{2}}\left[
\begin{array}{cc}
1 & 1 \\
-1 & 1
\end{array}
\right]
\]

\[
\mathbf{L_{1}}=\left[
\begin{array}{cc}
\cos{(\epsilon\cos{\omega_{1}t_2})} & \sin{(\epsilon\cos{\omega_{1}t_2})} \\
-\sin{(\epsilon\cos{\omega_{1}t_2})} & \cos{(\epsilon\cos{\omega_{1}t_2})}
\end{array}
\right]
\]

\[
\mathbf{L_{5}}=\left[
\begin{array}{cc}
\cos{(\epsilon\cos{\omega_{5}t_7})} & \sin{(\epsilon\cos{\omega_{5}t_7})} \\
-\sin{(\epsilon\cos{\omega_{5}t_7})} & \cos{(\epsilon\cos{\omega_{5}t_7})}
\end{array}
\right]
\]
and 
\[
\mathbf{L_{i}}=\left[
\begin{array}{cc}
\cos{(\epsilon\cos{\omega_{i}t_4})} & \sin{(\epsilon\cos{\omega_{i}t_4})} \\
-\sin{(\epsilon\cos{\omega_{i}t_4})} & \cos{(\epsilon\cos{\omega_{i}t_4})}
\end{array}
\right]
\]
for $i=2,3,4$.

\begin{figure}
\includegraphics[scale=1.0]{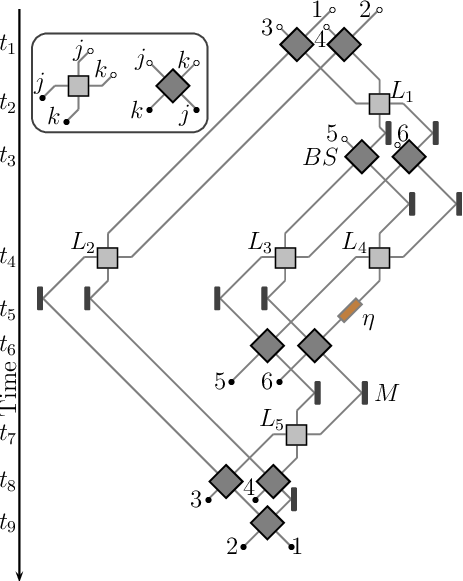}
\caption{
Six-port interferometer with empty dots showing the
input ports and filled  dots showing the output ports.  The
dark square boxes are the beam-splitters ($BS$),  the light
boxes are the time dependent $L$ elements and the long dark
rectangle are the mirrors.  The top left corner shows the
The input and output ports for $L$ and $BS$.}
\label{figure_1}
\end{figure}

For the purpose of possible experimental realization of the
sequence of unitaries on a state in six dimensional Hilbert
space, one can think of a single photon interferometer with
six-ports as detailed in Figure.~(\ref{figure_1}). In this
setup a single photon inside the interferometer can be in a
superposition of six non-overlapping ports forming a
six-dimensional Hilbert space. The single photon prepared
in superposition of being present in six ports at time $t_0<t_1$
undergoes the sequence of unitaries $U_1,U_2,\cdots,U_9$ at
moments of time $t_1,t_2,\cdots,t_9$ respectively. The first
two zero-loss beam-splitters (BS) having transmission and reflection
coefficients of one-third and two-third respectively act on
pairs of ports-$1,3$, and ports-$2,4$ to generate the unitary
time transformation $U_1$ at time $t_1$. $U_2$ is generated by
an element $L_1$, which is also a beam-splitter but with a time
varying reflectivity, acting on pair of port-$3$ and port-$4$.
The reflectivity of $L_1$ is so small that the probability
amplitude in either of the input ports ($3$ or $4$) remains
unaffected but at the same time it transfers a very tiny amplitude
between the ports in either direction as a leakage process so that
it can make a contribution in providing information about the past
of the photon in interferometer. The role of time dependence of
$L_1$ will be become clear from the discussion that
follows. The unitary $U_5$ is generated
by a phase shifter $\eta$ which produces a phase shift of $\pi$ in
the probability amplitude of photon in the port-$6$. Rest of the
unitaries can be easily related to processes presented in
Figure~(\ref{figure_1}) which are either BS or leakage processes.

As  we shall see, the \textit{leakage} processes described
above are engineered to provide us with a tool to
investigate the past of the particle.  A  \textit{leakage}
process between two completely empty boxes will definitely
not make any contribution to the time evolution of the state
of the particle and hence will not have any measurable
effects.  Therefore, the measurable effect  of such a
\textit{leakage} process, between any two boxes in the state
of the particle is an evidence that the amplitude of the
particle was not zero at least in one of the boxes involved
in the \textit{leakage} interaction.  It is easy to see that
due to the \textit{leakage} process, the change in the
probability amplitude of the particle being in one box is
proportional to the probability amplitude of it being in the
other box.

The initial state of the  particle is prepared in $\ket{1}$
(a single photon enters the interferometer through port-$1$) 
at time $t_{0}$. The probability of finding the particle in
state $\ket{1}$ (the photon in output port-$1$) at time
$t^\prime>t_{9}$, retaining only terms linear in $\epsilon$,
is calculated using Equation~(\ref{Hamiltonian_effective}),
according to Born rule, as:

\begin{equation} \label{final_prob1}
P=\frac{1}{18}\{1+2\epsilon(2\cos\omega_{1}t_2-
\cos\omega_{2}t_4+\cos\omega_{3}t_4+\cos\omega_{4}t_4)\}
\end{equation}

As one would expect, the probability $P$ depends on the reflectivity
of the time dependent beam-splitters $\{L_i\}$ at the moments of time
when the localized wave packet of photon passes through them.
Looking carefully at experimental setup shown in Figure~(\ref{figure_1}),
we can say $t_1,t_2,\cdots,t_9$ are not independent variables
present in unitaries but are dependent on the time when the photon
enters the interferometer. To make latter point clearer we emphasize 
the fact that the optical path-length of the photon travelling from one
optical device to another is fixed over time. In other words one can
say that time difference between the  unitaries is
fixed. Consider the optical path-length from source to
optical device generating $U_j$ to be $l_j$, then one has
$t_j=t_0+l_j/c$ where $t_0$ is the time when photon leaves
source. The time of detection of the photon at port-$1$  is $t^\prime$ 
and $t^\prime-t_0=\tau$ is constant across various
repeatitions of the experiment.
Physically it means that the photon takes $\tau$ time to travel
from source to detector each time the experiment is carried out.
Using the relation $t_j=t^\prime+( l_j/ c-\tau)$, 
Equation~(\ref{final_prob1}) can be re-written as a function
of $t^\prime$ as follows:
\begin{equation} \label{final_prob2}
\begin{split}
P=\frac{1}{18}[1+2\epsilon\{2\cos(\omega_{1}t^\prime-\theta_1)-
\cos(\omega_{2}t^\prime-\theta_2)
\\ 
+\cos(\omega_{3}t^\prime-\theta_3)+\cos(\omega_{4}t^\prime-\theta_4)\}]
\end{split}
\end{equation}
Where $\theta_1=\omega_{1}(\tau-l_1/c), \theta_i=
\omega_{i}(\tau-l_4/c)$ for $i=2,3,4$ depend on
oscillation frequencies of various 
time varying beam splitters and the geometry (optical 
path-length) of the interferometer and hence are constant phases.
Further assuming the condition $\omega_{j}^{-1}\gg{\tau}$ with 
$j=1,2,\cdots,5$ 
which
gives $\theta_{i}\ll{1}$ for $i=1,2,3,4$;
Equation~(\ref{final_prob2}) can be simplified as:
\begin{equation} \label{final_prob}
P\approx\frac{1}{18}[1+2\epsilon(2\cos\omega_{1}t^\prime-
\cos\omega_{2}t^\prime+\cos\omega_{3}t^\prime+\cos\omega_{4}t^\prime)]
\end{equation} 

Probability $P$ depends on reflectivities of various $L_j$'s
at the time when photon passes through them. 
Equations~(\ref{final_prob2}) and (\ref{final_prob})
are our main results. We use Equation~(\ref{final_prob})
in drawing operational inferences about the past of the
photon. It is to be emphasised here that these inferences
can be drawn by using Equation~\ref{final_prob2} also
to avoid any misunderstanding regarding the approximation
$\omega^{-1}_j\gg{\tau}$, however, Equation~\ref{final_prob}
is simpler and more convenient to use.

Experiments with a single particle cannot reveal any
information about the time dependency of probability $P$,
but experimental runs over ensembles with varying time can
provide information about the frequencies present in the
modulated probability $P$. As we
describe next, the experimental realisation of
Equation~(\ref{final_prob}) can be achieved if we sample a
sufficient number of photons in a  time window in which the
time dependent optical elements in the circuit do not vary
appreciably.
\subsection{Sampling protocol}
The probability $P$ can be experimentally measured by
repeating the experiment a large number of times at a certain
rate.  We need to have the frequencies $\omega_i$
sufficiently small 
so that we can measure over a
sufficiently large number of particles before the time
varying elements $L_i$ changes appreciably.
Suppose at each time $t=t_0+2n\tau$ where $n=0,1,2,\cdots,N_s$,
a particle is pre-selected which will
undergo a post-selection measurement at time $t=t^\prime+2n\tau$.
$N_s$ is the number of particles pre-selected
in one sample-run.
Note that not all pre-selected particles get post-selected.
A particle found in state $\ket{1}$ is counted,
otherwise it is  discarded. 
Right after each post-selection measurement
the experimental setup is kept ready to perform
pre-selection on a new particle.The sampling time period
$T_s=2\tau N_s$ is time taken to run experiment on a sample
of $N_s$ particles. $N_s$ and $\epsilon$ must be chosen in
such a way that the operational condition
$\epsilon^2{N_s}\approx{1}$ is satisfied.

As we shall see this can be easily achieved with photons. For a
precise measurement of modulations, the change in the number of
post-selected particles in each consecutive sample is
required to be smooth, hence $1\gg{T_s{\omega_i}}$ is
necessary. The number of post-selected particles in the $k^{th}$
sample is:
\begin{equation}
\label{Sampling}
\begin{aligned}
N_{k}={}&
\frac{N_s}{18}+\frac{\epsilon{N_s}}{9}[2\cos\{\omega_{1}(2k-1)\frac{T_s}
{2}\} \\ 
&-\cos\{\omega_{2}(2k-1)\frac{T_s}{2}\}+\cos\{\omega_{3}(2k-1)\frac{T_s}{2}\}
\\
& +\cos\{\omega_{4}(2k-1)\frac{T_s}{2}\}]
\end{aligned}
\end{equation}
Due to $\epsilon{N_s}\gg{1}$, the (co)sinusoidal
oscillations can be observed. The Fourier analysis of the
best fit of (\ref{Sampling}) reveals the frequencies
$\omega_i$.

In case of photon: $\tau=\frac{l}{c}$, here $l$ is the
optical path-length - the distance each photon travel
from source to detector in the
interferometer. The requirements for weakness of $\epsilon$
and sampling are: $\epsilon^{2}{N_s}\approx{1}$ and
$1\gg{T_s\omega_i}$. That gives
$\omega_i\ll\frac{c\epsilon^2}{2l}$ (here we have used
$N_s=\frac{T_s}{2\tau}$). For an interferometer of length
one meter and $\epsilon\approx{10^{-2}}$,
$\omega_i\ll{15000}$. The choice of $\omega_i\approx{100}$
is reasonable. For photons with well localized wave-packets,
one can increase $N_s$ (hence decrease $\epsilon$) by
sending a train of photons with a small spacing between
the two successive photons into the interferometer.
\subsection{Where was the photon?}
We make use of Equation~(\ref{final_prob}) to draw
inferences about the presence of the photon at various
locations inside the interferometer.  The
appearance of any observable signature of a localized-device
in the post-selection probability is considered as an indicator of the
presence of the particle at that location.  In an
experimental setup involving (co)sinusoidal time varying
\textit{leakage processes} $L_j$'s with various frequencies
$\omega_j$'s, our operational definition of the past says:
\textit{it cannot be possible that the particle
was not present at the location where $L_i$
is installed if frequency $\omega_i$ corresponding to device
$L_i$ is present in the modulated probability $P$ of
post-selection}. Therefore, we interpret the past using the following
principle:

\noindent 
S-$C$: \textit{A quantum particle cannot carry information
about a localized object without interacting with it. In
particular, if the particle is a photon inside an
interferometer, it cannot not visit the location of a
localized optical device and still gain  information about
it.}

Let us now look at Equation~(\ref{final_prob}) and draw
valid inferences about the past of a photon inside the
interferometer under discussion. Appearances of frequencies
$\omega_1,\omega_2, \omega_3$, and $\omega_4$ tell a story
about the past of the photon: one cannot say with certainty that
the photon, pre- and post-selected at entrance
and exit of port-$1$ respectively,
has not been at anyone or more of the locations
where time varying beam-splitters
$L_1, L_2, L_3$ and $L_4$ are installed.

The key result of this section to be emphasized for further
use is that one cannot claim with certainty that
the photon entered the interferometer through
port-$1$ and detected at output port-$1$ was not present at
$L_1$ at any intermediate time. 

\section{TSVF analysis of the gedanken experiment}
\label{TSVF}
Let us now explore the predictions of the 
TSVF of quantum mechanics for our gedanken experiment.
In order to answer the question whether the particle was
present in at least one box of the pair of boxes right
before the \textit{leakage} took place, we perform weak
measurements on both boxes. The weak traces present in
the pointer state after the post-selection will reveal the
presence of the particle. For the particle pre-selected in
state $\ket{\psi}=\ket{1}$ at time $t_0$ and post-selected
in the state $\ket{\phi}=\ket{1}$ at time $t^\prime$, we
calculate the weak values of the projection operators at
those boxes at the corresponding times. The weak value of
projection at box $k$ (port-$k$ of interferometer in
case of photon) right before time $t_j$ is written
$\Pi_{k}^{w}(t_{j})$. The weak values of projections $\Pi_k
=\ket{k}\bra{k}$ right before all the
\textit{leakage} processes come out to be:
\begin{eqnarray}
	\Pi_{3}^{w}(t_{2}) & = & \epsilon(2\cos{\omega_{1}t^\prime}+
	\cos{\omega_{3}t^\prime}+\cos{\omega_{4}t^\prime})
	\nonumber 
	\\
	\Pi_{1}^{w}(t_{4}) & = & 1-\epsilon(2\cos{\omega_{1}t^\prime}+
	\cos{\omega_{3}t^\prime}+\cos{\omega_{4}t^\prime})
	\nonumber 
	\\
	\Pi_{3}^{w}(t_{4}) & = & -1+\epsilon(2\cos{\omega_{1}t^\prime}-
	\cos{\omega_{2}t^\prime}+2\cos{\omega_{3}t^\prime}
	+\cos{\omega_{4}t^\prime}+\cos{\omega_{5}t^\prime})
	\nonumber 
	\\
	\Pi_{4}^{w}(t_{4}) & = & \epsilon\cos{\omega_{1}t^\prime} 
	\nonumber 
	\\
	\Pi_{5}^{w}(t_{4}) & = & 1-\epsilon(2\cos{\omega_{1}t^\prime}-
	\cos{\omega_{2}t^\prime}+\cos{\omega_{3}t^\prime} 
	+\cos{\omega_{5}t^\prime})
	\nonumber 
	\\
	\Pi_{6}^{w}(t_{4}) & = & \epsilon\cos{\omega_{1}t^\prime}
	\nonumber 
	\\
	\Pi_{4}^{w}(t_{7}) & = & \epsilon(2\cos{\omega_{1}t^\prime+
		\cos{\omega_{3}t^\prime}+\cos{\omega_{4}t^\prime}})
	\nonumber
	\\
	\Pi_{3}^{w}(t_{7})  &=& 
	\Pi_{4}^{w}(t_{2})  =   
	\Pi_{2}^{w}(t_{4})  =  0 
	\label{weak_values}
\end{eqnarray}

\begin{figure}
	\begin{center}
		\includegraphics[scale=1]{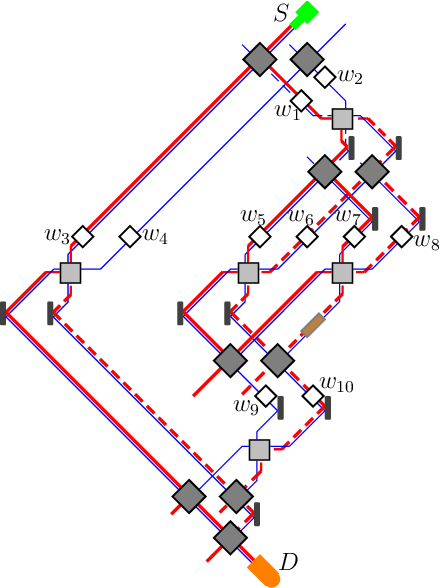}
		\caption{
			The thick (red) and thin (blue) lines represent the forward and
			backward evolving state vectors of the single photon, pre
			nd post-selected at the source $S$ and detector $D$,
			respectively. The solid lines represent the non vanishing
			and significant probability amplitude, dashed lines
			represent insignificant (order $\epsilon$) probability
			amplitudes, and the absence of a line represents amplitudes
			which are zero or proportional to higher powers of
			$\epsilon$. $w_1,w_2,\cdots,w_{10}$ denote weak 
			measurement devices of corresponding projection operators.}
		\label{figure_2}
	\end{center}
\end{figure} 
\newpage
\newpage
\subsection{Measurement of weak values}
The weak values can be measured by
introducing weak von Neumann
type interaction between the system and the pointer with
interaction Hamiltonian between the system and the apparatus given by
\begin{equation}
	H_{\rm SA} = \kappa \delta(t-t^\prime) \hat{A} \otimes \hat{p}
	\label{weak_hamiltonian}
\end{equation}
Where $\kappa$ is the strength of the measurement, $\hat{A}$
is the observable being measured (in our case it is the
projection operator onto a particular location) and
$\hat{p}$ is the pointer momentum operator.  
The measurement is weak when $\kappa \ll 1$. After this interaction,
the displacement of the pointer state vector is proportional to the
weak value of the observable being measured. The initial
state of the ancillary system is taken to be a Gaussian with
a finite width. In the case of photon, one can use the frequency
space of the photon as a pointer and perform weak coupling using
electro-optics phase modulators (EOM)~\cite{PhysRevA.95.042121}.
The weak interaction leads to a small shift in the center of
the Gaussian state, which is the measure of the weak trace that
the photon leaves on the ancillary system.

The experimenter in a weak measurement scenario has complete
control over the size of pre-and post selected ensemble,
state of the pointer and the weak measurement interaction strength;
and can tune these parameters suitably so that weak
values can be measured up to a desired precision.
Weak nature of the measurement implies that the effects
of higher powers of coupling strength $\kappa$ are not
recordable experimentally. In an ideal weak measurement scenario,
the choice of $\kappa$ and the size of the ensemble $N$ should be
such that $N\kappa^2\rightarrow{1}$ when $N\rightarrow{\infty}$.
The ideal condition $N\rightarrow{\infty}$ is not feasible, therefore,
the experimenter can choose $\kappa$ and $N<\infty$ such that
$N\kappa^2\approx{1}$ while $0<\kappa\ll 1$ in all
practical scenarios.

\subsection{Where was the photon according to TSVF?}
The story told by weak values is surprisingly different.
For a single photon, pre-selected
in input port-$1$ and post selected in output port-$1$,
the weak values of projection operators at
locations of weak measurements $w_1,w_2,\cdots,w_{10}$ shown
in Figure~(\ref{figure_2}) are detailed in
Equation~(\ref{weak_values}). The values reveal that the
presence of the particle was of the order of
$1$ at $w_3,w_5,$ and $w_7$ and of the order of first or
higher powers of $\epsilon$ at the rest of the locations.
Particularly, for port-$3$ and port-$4$, between $t_1$ and
$t_2$, at least one of the forward and backward evolving
wave-functions vanishes to order $\epsilon$ (see
Figure~(\ref{figure_2}) for pictorial representation).
To see contradiction between the conclusion drawn
in subsection~\ref{4.2} and the prediction (retrodiction) of TSVF,
let us consider following two cases:

\paragraph*{Case 1:} The parameter $\epsilon$ of the interferometer
is tuned in such a way that $N\epsilon^2\rightarrow{1}$ when
$N\rightarrow{\infty}$. If experiment (as described in section~\ref{gedanken})
is performed on infinitely large ensemble ($N\rightarrow\infty$),
there will be no traces of $\epsilon^2$ or higher orders in
the final probability but at the same time one can record deviations of
the order $\epsilon$. Once the pre-and post-selected ensemble (which
is defined by pre-and post-selection states and all the unitaries including
time varying beam-splitters $\L_i$) is fixed, the experimenter can deploy
weak measurement schemes to investigate past of the photons according
to TSVF. The most optimal weak measurement setup requires
$N\kappa^2\rightarrow{1}$ when $N\rightarrow{\infty}$. This amounts
to $\kappa^2 \approx{0}$ and we already have $\epsilon^2\approx{0}$,
therefore, we conclude that $\kappa\epsilon\approx{0}$,
which implies that the weak traces corresponding to weak
values of the order $\epsilon$ are too small to be observed (even ideally) in this case.
In other words \textit{operational condition},
$N\epsilon^2\rightarrow{1}$ and $N\kappa^2\rightarrow{1}$ when
$N\rightarrow{\infty}$, implies $N\kappa\epsilon\rightarrow{1}$.
Since the weak values of order $\epsilon$ (in this case) are not
experimentally measurable, according to Equation~(\ref{weak_values});
the photon leaves weak traces only at ports
$1, 3$ and $5$ with nonzero weak values
$\Pi_{1}^{w}(t_{4}),\Pi_{3}^{w}(t_{4})$, and $\Pi_{5}^{w}(t_{4})$
respectively. The information about the
presence of the photon in the pair of ports $3$ and $4$ just
before $L_1$ is completely absent from the weak signal, which leads
us to draw a conclusion on the basis of weak value based
operational definition of the past of a quantum particle:
{\it the photon has not been in the vicinity of time
	varying beam-splitter $L_1$.} This prediction is in direct contradiction with our
earlier conclusions based on standard quantum mechanical
analysis under same approximations.

\paragraph*{Case 2:} Consider a case where the parameter $\epsilon$ of the
interferometer is tuned in such a way that $N\epsilon^2\approx 1$
for some finite $N$ and $0<\epsilon\ll 1$. Under these conditions,
the experimenter can choose arbitrary large ensemble $N^\prime\gg N$
in weak measurement setup and can easily record weak values of order
$\epsilon$. Now, for time being, imagine a situation where experimenter
chooses to perform experiment with $N$ systems. Although, this is not an
optimal weak measurement setup but one can draw certain
inferences based on TSVF retrodiction using ABL rule. As we have discussed
earlier, TSVF goes hand in hand with ABL rule. ABL rule can be expressed
in terms of weak values, using equations~(\ref{ABL_rule_TSVF_degenerate})
and~(\ref{WV}), as:

\begin{equation}
	\label{ABL_rule_WV}
	P_{t}(a_n|\psi_1,\psi_2)=\frac{|\Pi_{a_n}^{w}|^{2}}
	{\sum_{i}{|\Pi_{a_i}^w|^{2}}}
\end{equation}
Now we ask the following question: given that the experimenter performs
experiment on finite number of systems
$N$ such that $N\epsilon^2\approx 1$, how many
systems would have been found in if the box $i$ were opened at some
intermediate time $t$? The answer, according to ABL rule, is $NP_{t}(i)$,
where
\begin{equation}
	\label{ABL_rule_int}
	P_{t}(i)=\frac{|\Pi_{i}^{w}(t)|^{2}}
	{|{(I-\Pi_{i})}^{w}(t)|^{2}+|\Pi_{i}^{w}(t)|^{2}}.
\end{equation}
Weak values $\{\Pi_{i}^{w}(t)\}$ presented in equation~\ref*{weak_values}
dictate us to conclude that less than one out of $N$ systems would
have been found in box $3$ and $4$ at $t_2$ (in ports $3$ and $4$
just before $L_1$) if the respective boxes were opened. Which
leads us further to conclude that no photon was present in vicinity of
$L_i$ if ensemble size was $N$. On the other hand,
equation~(\ref{final_prob}) (more explicitly equation~(\ref{Sampling})
suggests that given ensemble of $N$ photons or a significant
fraction of it (at least of order $\epsilon N$) does carry 
information about the time varying element $L_1$. In light of S-$C$,
one can safely conclude that one cannot claim with certainty that
out of the $N$ photons which entered the interferometer through port-1 and
were detected at output port-1 no (or only at most of order one) photons
were present at $L_1$.

\subsection{CH Analysis of the Gedanken Experiment}
\label{CH_analysis}
Along with TSVF approach, Consistent Histories (CH) formalism
has also been used to analyze the past of a quantum particle.
Therefore, we consider the CH formalism for our gandanken
experiment.
In CH formalism, one can talk about the trajectories of a
quantum system only when the conditions of 
\textit{single framework} and \textit{consistency} are
satisfied. A family of consistent histories represents the
past of a quantum system. CH formalism considers inconsistent
histories of a quantum system `meaningless'. To analyze
the past of a photon in our interferometer using CH formalism,
let us consider the family of consistent histories (see ref.
\cite{PhysRevA.94.032115}for notations):
\begin{equation}
	\label{family of CH}
	\mathcal{F}:\Pi^{(t_1)}_1\odot\{\Pi^{(t^\prime)}_1,\Pi^{(t^\prime)}_2,\Pi^{(t^\prime)}_3,\Pi^{(t^\prime)}_4,\Pi^{(t^\prime)}_5,
	\Pi^{(t^\prime)}_6\}
\end{equation}
Here $\Pi_k=|k\rangle\langle{k}|$ and $\Pi^{(t)}_k$
represents the physical property that `photon is in
$|k\rangle$ at time $t$'.
$\mathcal{F}$ can be divided into two families as per our interest:
\begin{equation}
	\label{family1}
	\mathcal{F}_1:\Pi^{(t_1)}_1\odot\Pi^{(t^\prime)}_1=\Pi^{(t_1)}_1\odot{I}^{(t_2)}\odot{I}^{(t_3)}\cdots\odot{I}^{(t_9)}\Pi^{(t^\prime)}_1
\end{equation}
and
\begin{equation}
	\label{family2}
	\mathcal{F}_2:\Pi^{(t_1)}_1\odot\{\Pi^{(t^\prime)}_2,\Pi^{(t^\prime)}_3,\Pi^{(t^\prime)}_4,\Pi^{(t^\prime)}_5,
	\Pi^{(t^\prime)}_6\}
\end{equation}

Only consistent histories of 
family $\mathcal{F}_1$ are of our
interest in which photon enters port-1
and detected in port-1. 
One can easily verify that family
$\mathcal{F}_1$ can only be refined 
further if all $L$ elements are
identities simultaneously, which is
possible only if $\epsilon=0$ or
$\cos{\omega_1{t}}=\cos{\omega_2{t}}
=\cos{\omega_3{t}}=\cos{\omega_4{t}}
=\cos{\omega_5{t}}=0$
which is momentary during
the experiment run. In that case photon
remains in port-1 throughout the time
according to CH formalism. Since 
$\epsilon\neq{0}$ in the case of our gadenken experiment,
CH formalism provide no information
about the past of the photon other than saying that there is
no consistent history corresponding to the trajectory being
considered.
\section{Conclusions and Discussion}
\label{conc}
Truthfulness of S-$C$ asserts that
it cannot be the case that the photon
did not pass through $L_1$
with certainty while S-$B$ asserts that
it did not have a passage through $L_1$
with certainty given that the operational
condition $N\kappa^2\approx{1}$ and $N\epsilon^2\approx{1}$ with
$N\gg{1}$ is satisfied. Even when the operational condition
is not satisfied, a clear difference in the quantitative
presence of photon inside the interferometer at various locations
can be seen in the two different approaches. For instance,
quantification of presence in TSVF is in terms of weak values of
the position projection operators, according to which presence
of photon near $L_1$ is very smaller than those of near
$L_2, L_3$ and $L_4$; while any possible quantification
of the presence based on amplitudes of oscillating
terms present in equation~(\ref{final_prob})
suggests it was of an equal order.

In the language of counterfactual ABL rule, less than
one (which is zero) photons would have been detected if one
had tried to detect $N$ pre- and post selected photons in
entrance ports of $L_1$ indicating
no presence of a photon near $L_1$. The contradictory
conclusions inferred from two assertions imply: \textit{at
	least one of S-$C$ and S-$B$ is false.} Since S-$C$ is based
on the fact that all the interactions in nature are local
and the operational definition of past based on weak values
itself is implicitly based on S-$C$, one is
forced to forgo
S-$B$. This
further leads us to conclude that the S-$A$ is false \ie if
weak value of a projection operator $\ket{x}\bra{x}$ is
zero, then it is not necessary that the particle is not
present at location $x$. This invalidates the WVH that the
weak value of an observable is the value of that observable
{\ie} if weak value is zero then the system does not carry
the corresponding
property.

As we have seen in section~\ref{paradoxes}, all weak value (TSVF) paradoxes
are based on correctness and rationality of WVH more specifically
truthfulness of S-$A$, our results therefore, have a bearing on these
paradoxes. As per our conclusions,
absence of certain traces in Danan \etal experiment does not imply
that photon does not pass through those regions taking discontinuous
trajectories to reach the detector. Similarly, zero weak values of
certain observables does not imply circular polarization of a photon
is separated from the wave function in quantum Cheshire cat paradox.
Same is applicable to weak value version of Hardy's paradox.

A natural question arises: what are weak values if not
properties of systems? What do weak values tell about the
properties of systems between two successive measurements? A
plausible answer is given by D.
Sokolovski~\cite{SOKOLOVSKI20161593}: weak value of an
observable is the transformation generated by weak
measurement unitaries on pre-selected state which reaches
the post-selection. If the observable is projection operator
$\ket{a}\bra{a}$ then the weak value is relative transition
amplitude of pre-selected state $\ket{\psi}$ to post-selected
state $\ket{\phi}$ through state $\ket{a}$.

We have shown in our analysis that the conclusions based on
TSVF and the related weak values (WVH) may not always be
correct in the context of the past of a quantum system.  We
clearly obtain nonzero signals in our measurable probability
distribution from the regions of the interferometer where
TSVF claims that the photon never entered or the presence of
photon was not measureable.  It will be interesting to
explore the possibilities of carrying out such
interferometric experiments. More detailed investigations
are required to pin down the exact role of weak traces and
circumstances where they play a significant role in
providing information about the particle trajectories.

\begin{acknowledgments}
Arvind acknowledges the financial
support from {\sf DST/ICPS/QuST/Theme-1/2019/General}
Project number {\sf Q-68}.
\end{acknowledgments}

%

\end{document}